\documentclass[conference]{IEEEtran}
\IEEEoverridecommandlockouts
\usepackage{cite}
\usepackage{amsmath,amssymb,amsfonts}
\usepackage{algorithmic}
\usepackage{graphicx}
\usepackage{textcomp}
\usepackage{listings}
\usepackage{braket}
\usepackage{enumitem}

\def\BibTeX{{\rm B\kern-.05em{\sc i\kern-.025em b}\kern-.08em
    T\kern-.1667em\lower.7ex\hbox{E}\kern-.125emX}}
\begin{document}

\title{GPU-Accelerated Selected Basis Diagonalization with Thrust for SQD-based Algorithms\\

\thanks{Identify applicable funding agency here. If none, delete this.}
}

\author{
\IEEEauthorblockN{Jun Doi}
\IEEEauthorblockA{\textit{IBM Quantum} \\
\textit{IBM Research - Tokyo}\\
Tokyo, Japan \\
doichan@jp.ibm.com}
\and
\IEEEauthorblockN{Tomonori Shirakawa}
\IEEEauthorblockA{\textit{Center for Computational Science} \\
\textit{RIKEN}\\
Kobe, Japan \\
t-shirakawa@riken.com}
\and
\IEEEauthorblockN{Yukio Kawashima}
\IEEEauthorblockA{\textit{IBM Quantum} \\
\textit{IBM Research - Tokyo}\\
Tokyo Japan \\
yukio.kawashima@ibm.com}
\and
\IEEEauthorblockN{Seiji Yunoki}
\IEEEauthorblockA{\textit{Center for Computational Science} \\
\textit{RIKEN} \\
Kobe, Japan \\
yunoki@riken.jp}
\and
\IEEEauthorblockN{Hiroshi Horii}
\IEEEauthorblockA{\textit{IBM Quantum} \\
\textit{IBM Research - Tokyo}\\
Tokyo Japan \\
horii@jp.ibm.com}
}

\maketitle

\begin{abstract}
Selected Basis Diagonalization (SBD) plays a central role in Sample-based Quantum Diagonalization (SQD), where iterative diagonalization of the Hamiltonian in selected configuration subspaces forms the dominant classical workload. We present a GPU-accelerated implementation of SBD using the Thrust library. By restructuring key components---including configuration processing, excitation generation, and matrix-vector operations---around fine-grained data-parallel primitives and flattened GPU-friendly data layouts, the proposed approach efficiently exploits modern GPU architectures. In our experiments, the Thrust-based SBD achieves up to $\sim$40$\times$ speedup over CPU execution and substantially reduces the total runtime of SQD iterations. These results demonstrate that GPU-native parallel primitives provide a simple, portable, and high-performance foundation for accelerating SQD-based quantum--classical workflows.
\end{abstract}


\section{Introduction}

Hybrid quantum--classical algorithms have emerged as a powerful approach for studying strongly correlated quantum systems, in which quantum processors generate physically meaningful samples while classical high-performance computing (HPC) resources perform the demanding post-processing required for accurate energy estimation. Among these approaches, \emph{Sample-based Quantum Diagonalization} (SQD)~\cite{robledo2024chemistry} has recently been proposed as a workflow that iteratively refines a subspace of many-electron configurations based on samples obtained from a quantum circuit. By encoding a molecular Hamiltonian into a parameterized quantum circuit and repeatedly sampling the resulting quantum state, SQD obtains electronic configurations whose probabilities reflect their importance for the target ground or excited states. These configurations are then filtered and ranked on classical hardware to construct a reduced Hilbert-space subspace on which a diagonalization procedure is performed.

A defining characteristic of SQD is that the \emph{classical diagonalization step dominates the overall computational cost}, particularly for chemically accurate ground-state calculations. Each SQD iteration requires solving an eigenvalue problem on a selected configuration space that may contain up to $10^8$--$10^{10}$ determinants. Iterative eigensolvers such as the Davidson method therefore spend most of their runtime evaluating Hamiltonian matrix--vector products, whose cost scales with the size of the reduced basis. As a result, large-scale SQD calculations require substantial HPC resources even when quantum sampling is efficient.

Recent large-scale SQD studies ~\cite{robledo2024chemistry, shirakawa25} have been performed on the Fugaku supercomputer using a highly optimized CPU-based implementation of Selected Basis Diagonalization (SBD) ~\cite{sbd}. These studies demonstrated that SQD can scale to configuration spaces far beyond the reach of exact diagonalization when supported by massive CPU parallelism and large memory footprints. However, modern leadership-class supercomputers increasingly adopt GPU accelerators as their primary compute engines, as seen in systems such as Frontier, Aurora, and other exascale platforms. To take full advantage of these architectures, diagonalization methods must be redesigned to exploit fine-grained parallelism, high thread concurrency, and GPU-resident data structures.


Furthermore, as SQD evolves, the classical diagonalization backend must support a broader range of Hamiltonian evaluation strategies.
While the tensor-based ``half-bitstring'' representation adopted in prior work is widely used and effective for certain structured systems, our implementation is designed to \emph{generalize} this approach to also support full-bitstring representations, arbitrary spin symmetries, and more flexible operator forms.
Although full-bitstring support is not yet enabled in the current release, the data layouts and excitation iterators have been structured to admit this extension with minimal changes.
These requirements further motivate the development of GPU-native data layouts and parallel algorithms capable of sustaining high performance for both structured and sparse Hamiltonians.

In this work, we present a GPU-accelerated implementation of Selected Basis Diagonalization tailored for SQD-based algorithms. Unlike directive-based approaches, our method utilizes the Thrust parallel algorithms library to construct a fully GPU-native SBD backend, enabling fine-grained control over memory access patterns, data-parallel execution, and persistent device-resident computation. By flattening configuration data structures, restructuring excitation evaluation, and implementing all performance-critical components using GPU-centric primitives, the proposed method achieves substantial speedups on modern GPU clusters.

The resulting GPU-optimized SBD supports both half-bitstring and full-bitstring representations within a unified framework and provides an efficient classical backend for next-generation SQD workflows. As GPU-based HPC systems continue to dominate the landscape, such GPU-specialized diagonalization capabilities will be essential to scale hybrid quantum--classical algorithms to chemically meaningful system sizes.

\section{Classical Diagonalization in SQD: Problem Structure and Computational Challenges}
\label{sec:diag-overview}

In Sample-Based Quantum Diagonalization (SQD), repeated quantum measurements identify a subset of many-electron configurations that are most relevant to a target state (typically the ground state). The classical component then constructs an effective Hamiltonian on this selected subspace and solves a reduced eigenproblem to obtain approximate eigenpairs. Let $\{\Phi_i\}_{i=1}^{N}$ denote the selected configurations (Slater determinants) and let
\begin{equation}
  H_{ij} \;=\; \braket{\Phi_i | \hat{H} | \Phi_j}, \qquad
  \mathbf{H} \in \mathbb{R}^{N \times N},
\end{equation}
be the Hamiltonian restricted to that subspace. The computational objective is to extract the lowest few eigenpairs
\begin{equation}
  \mathbf{H}\,\mathbf{c} \;=\; E\,\mathbf{c},
\end{equation}
with emphasis on the smallest eigenvalue $E_0$ (ground state) and its eigenvector $\mathbf{c}_0$.

\subsection{Matrix-Free vs.\ Explicit Assembly}
Two realizations are commonly employed:
\begin{enumerate}
  \item \textbf{Matrix-free (on-the-fly)}: The matrix is never formed explicitly; instead, only the action $\mathbf{y} \leftarrow \mathbf{H}\,\mathbf{x}$ is evaluated when required by the iterative eigensolver~\cite{knowles}. This reduces memory proportional to $N^2$ but shifts cost to repeated evaluation of $H_{ij}$ contributions.
  \item \textbf{Explicit (assembled)}: The nonzeros (often highly structured due to Slater--Condon rules) are precomputed and stored, enabling faster SpMV thereafter at the expense of memory.
\end{enumerate}
In large SQD problems ($N \gtrsim 10^8$), the matrix-free strategy is preferred due to memory constraints. Our implementation supports both, with performance emphasis on the matrix-free path.

\subsection{Hamiltonian Application and Excitation Structure}
Let the electronic Hamiltonian be written in second quantization
\begin{equation}
\hat{H} \;=\; \sum_{pq} h_{pq}\,\hat{a}^\dagger_p \hat{a}_q
\;+\; \frac{1}{2} \sum_{pqrs} (pq|rs)\,\hat{a}^\dagger_p \hat{a}^\dagger_q \hat{a}_s \hat{a}_r,
\end{equation}
with one- and two-electron integrals $(h_{pq})$ and $(pq|rs)$. By the Slater--Condon rules, $\braket{\Phi_i | \hat{H} | \Phi_j}$ is nonzero only when $\Phi_i$ and $\Phi_j$ differ by at most two spin-orbital occupations (``0-, 1-, or 2-electron excitations''). Consequently, the Hamiltonian action
\begin{equation}
  y_i \;=\; (\mathbf{H}\mathbf{x})_i \;=\;
  \sum_{j \in \mathcal{N}(i)} H_{ij}\,x_j
\end{equation}
reduces to iterating over the sparse excitation neighborhood $\mathcal{N}(i)$ of each configuration $i$.

\paragraph*{Half-bitstring vs.\ Full-bitstring.}
Prior work has exploited a tensor-factorized (``half-bitstring'') encoding that separates $\alpha$ and $\beta$ spin sectors, enabling structured traversal of allowed excitations. To broaden applicability (arbitrary spin structures, general operator forms, and future extensions), we also support a \emph{full-bitstring} representation in which a configuration is a single packed bitstring over all spin-orbitals. Both paths ultimately generate the same excitation sets $\mathcal{N}(i)$ and phase factors, but their data access patterns and combinatorics differ. Our diagonalization layer abstracts over the representation, exposing a uniform \texttt{apply\_H(x)$\to$y} interface.

\subsection{Iterative Diagonalization via Davidson}
We employ the Davidson method ~\cite{davidson} to compute the lowest eigenpairs of $\mathbf{H}$. Davidson is well suited to large, sparse, or matrix-free problems because it requires only (i) matrix--vector products, (ii) orthogonalization, (iii) a small projected eigenproblem, and (iv) a simple preconditioner. The basic iteration (for one target eigenpair) proceeds as follows:
\begin{enumerate}
  \item \textbf{Initialization:} Choose a normalized initial vector $\mathbf{v}_1$ (e.g., selected from the sampling distribution). Set the search subspace $\mathcal{V}_1 = \mathrm{span}\{\mathbf{v}_1\}$.
  \item \textbf{Subspace expansion:} For $k=1,2,\dots$
    \begin{enumerate}
      \item Compute $\mathbf{w}_j = \mathbf{H}\,\mathbf{v}_j$ for all basis vectors $\{\mathbf{v}_j\}_{j=1}^{k}$.
      \item Form the $k\times k$ projected matrix $\mathbf{T}_k = \mathbf{V}_k^\top \mathbf{W}_k$ with $\mathbf{V}_k=[\mathbf{v}_1,\dots,\mathbf{v}_k]$ and $\mathbf{W}_k=[\mathbf{w}_1,\dots,\mathbf{w}_k]$.
      \item Solve $\mathbf{T}_k \mathbf{y} = \theta \mathbf{y}$ for the smallest Ritz value $\theta$ and vector $\mathbf{y}$.
      \item Form the Ritz vector $\mathbf{u} = \mathbf{V}_k \mathbf{y}$ and residual $\mathbf{r} = \mathbf{H}\mathbf{u} - \theta \mathbf{u}$.
      \item \textbf{Convergence test:} If $\|\mathbf{r}\|_2 \le \varepsilon$ (target tolerance), accept $\theta$ and $\mathbf{u}$.
      \item \textbf{Preconditioning:} Construct the correction vector
      \begin{equation}
        \mathbf{t} \;\approx\; (\mathbf{D}-\theta \mathbf{I})^{-1}\mathbf{r},
      \end{equation}
      where $\mathbf{D}$ is a diagonal (or block-diagonal) approximation of $\mathbf{H}$ (e.g., diagonal of $\mathbf{H}$ in the selected basis). In practice, we cap or shift small denominators to ensure numerical stability.
      \item \textbf{Orthogonalization:} Orthogonalize $\mathbf{t}$ against $\mathcal{V}_k$ (Modified Gram--Schmidt with optional re-orthogonalization), normalize to obtain $\mathbf{v}_{k+1}$, and set $\mathcal{V}_{k+1}=\mathcal{V}_k \cup \{\mathbf{v}_{k+1}\}$.
      \item Optionally \textbf{restart} (thick-restart or windowed variants) to control the subspace dimension.
    \end{enumerate}
\end{enumerate}
For multiple eigenpairs, block-Davidson variants update several trial vectors per iteration and use block orthogonalization and a block-projected eigenproblem.

\subsection{Cost Drivers and Data Access Patterns}
Across iterations, the dominant cost is typically the Hamiltonian application $\mathbf{w}=\mathbf{H}\mathbf{v}$, which entails:
\begin{itemize}
  \item Traversal of the excitation neighborhood $\mathcal{N}(i)$ for each determinant $i$;
  \item Evaluation of $H_{ij}$ contributions (integral lookups and phase/parity from fermionic permutations);
  \item Accumulation $y_i \mathrel{+}= H_{ij} x_j$ with careful handling of numerical stability and parallel reductions.
\end{itemize}
The complexity scales with the total number of contributing excitation pairs, which, for electronic-structure Hamiltonians, is $O(N \cdot \bar{z})$ where $\bar{z}$ is the average number of connected determinants per row under the Slater--Condon rules. Orthogonalization and small dense eigenproblems in the projected space are lower-order costs as long as the subspace dimension remains modest (tens to low hundreds).

\subsection{Diagonal and Preconditioner Construction}
The Davidson preconditioner uses an approximation to $(\mathbf{H}-\theta \mathbf{I})^{-1}$, typically the inverse of the diagonal of $\mathbf{H}$ in the selected basis:
\begin{equation}
  M^{-1} \;\approx\; \mathrm{diag}\!\left(\frac{1}{H_{ii}-\theta}\right).
\end{equation}
Diagonal elements $H_{ii}$ are inexpensive to evaluate and can be cached. In practice we apply damped denominators,
\begin{equation}
  \frac{1}{H_{ii}-\theta} \;\leftarrow\; \frac{1}{\mathrm{sign}(H_{ii}-\theta)\,\max(|H_{ii}-\theta|,\delta)},
\end{equation}
with a small $\delta>0$ to avoid instabilities.

\subsection{Support for Half- and Full-Bitstring Paths}
Our diagonalization layer accommodates both excitation-generation schemes:

\begin{description}
  \item[Half-bitstring:] 
  Separate $\alpha$/$\beta$ bitstrings and structured single/double excitations per spin sector; efficient for systems with clear spin partitioning and favorable tensor layouts.

  \item[Full-bitstring:] 
  A single packed bitstring over all spin-orbitals; uniform treatment of spin/multireference patterns and straightforward extension to generalized operators. Bit operations such as population counts, bit scans, and parity dominate the arithmetic.
\end{description}

The \texttt{apply\_H} routine exposes a common interface while selecting the appropriate excitation iterator and integral-access backend.

\subsection{Stopping Criteria and Accuracy}
We monitor (i) the residual norm $\|\mathbf{r}\|_2$, (ii) the Rayleigh quotient stability $|\theta^{(k)}-\theta^{(k-1)}|$, and (iii) orthogonality loss in the subspace (via $\|\mathbf{V}_k^\top \mathbf{V}_k - \mathbf{I}\|_F$). Typical tolerances for ground-state energy are $10^{-8}$ to $10^{-10}$ in relative terms, with tighter criteria applied when subsequent SQD iterations depend sensitively on energy differences.

\subsection{Parallelization Opportunities}
The algorithm naturally exposes multiple levels of parallelism:
\begin{itemize}
  \item \textbf{Across configurations:} Rows $i$ are independent in $\mathbf{H}\mathbf{x}$.
  \item \textbf{Within a configuration:} Excitations in $\mathcal{N}(i)$ can be traversed in parallel with reductions on $y_i$.
  \item \textbf{Across vectors (block methods):} Batched Hamiltonian applications and orthogonalization for multiple trial vectors.
  \item \textbf{Distributed memory:} Domain decomposition over configuration indices, with halo-free matrix-free application and collective operations only for orthogonalization/projection.
\end{itemize}
These properties motivate GPU-specialized implementations discussed in later sections, where we employ data-parallel primitives and GPU-resident data to minimize synchronization and data movement.

\section{GPU Accelerated Diagonalization for Half-Bitstrings}
\label{sec:gpu-accel-diag}

In this section, we describe our GPU-oriented redesign of the diagonalization backend for half-bitstrings, used in SQD-based workflows. The dominant computational cost arises from the matrix-free evaluation of $\mathbf{y}=\mathbf{H}\mathbf{x}$ within the Davidson algorithm, which offers substantial thread-level parallelism through excitation enumeration under the Slater--Condon rules. To efficiently exploit modern GPU architectures, we adopt a data-parallel approach based on the Thrust C++ library, enabling STL-like programming while executing all performance-critical operations on the device. Our implementation supports both half-bitstring and full-bitstring representations, employs determinant caching to reduce redundant computation, and collapses excitation loops to expose large parallel index spaces. The subsections below describe our use of Thrust, the threading strategy for sparse matrix--vector multiplication, determinant caching, object-based kernel organization, and communication--computation overlap for distributed execution.

\subsection{Using Thrust C++ Library}
We used Thrust C++ library \cite{thrust} to implement GPU accelerated sparse matrix multiplication kernels for diagonalization. Thrust provides GPU-optimized memory abstractions and parallel algorithms through an interface closely aligned with the C++ Standard Template Library (STL), enabling rapid development of GPU kernels without low-level CUDA boilerplate. The GPU memory can be managed by \verb+thrust::device_vector+ class that has the same interface to \verb+std::vector+ class. So we can simply copy data between \verb+std::vector+ and \verb+thrust::device_vector+ by \verb+=+ operator. Thrust also provides powerful methods for accelerated linear algebras for \verb+thrust::device_vector+. In addition, we can apply custom GPU kernels to \verb+thrust::device_vector+ and code in C++ class, this helps to improve productivity and readability of GPU programming. The C++ code in Listing \ref{list_axpy} shows an example of custom GPU kernel program that calculates $a * X + Y$ for vector \verb+X+ and \verb+Y+. 

\begin{lstlisting}[float,caption={Thrust custom kernel class example for AXPY}, label={list_axpy}, language=c++,breaklines=true,columns=fixed,basewidth=0.5em]
template <typename T>
class AXPY_kernel {
protected:
    T a;
public:
    AXPY_kernel(T a_in) : a(a_in) {}

    __host__ __device__ T operator()(const T& x, const T& y) const
    {
        return a * x + y;
    }
};
\end{lstlisting}

Then \verb+thrust::transform+ function executes $a * X + Y$ kernel on the GPU as following code. \verb+thrust::transform+ executes a binary operation for each two vector elements and stores in the output vector.

\begin{lstlisting}[float,caption={Running a custom kernel with transform}, label={list_exec_axpy}, language=c++,breaklines=true,columns=fixed,basewidth=0.5em]
thrust::device_vector<double> X(N);
thrust::device_vector<double> Y(N);
double a;
// This calculates Y = a * X + Y
thrust::transform(thrust::device, X.begin(), X.end(), Y.begin(), Y.begin(), AXPY_kernel<T>(a));
\end{lstlisting}

For more general or complicated GPU kernels, \verb+thrust::for_each+ function provides indices of one dimensional loop for each GPU thread. 

Applications ported by using Thrust can be executed both on GPU and CPU, on the CPU GPU kernels are parallelized by OpenMP to run on CPUs. Thrust is basically designed for NVIDIA's GPUs because Thrust is included in CUDA Toolkit, but the code can be ported to AMD's GPUs by a little effort.

\subsection{Threading Sparse Matrix Multiplication} 
The \verb+mult+ routine computes sparse matrix-vector multiplication in three different tasks for each \verb+alpha*beta+ determinants space. Where three task types are dictated by the Slater-Condon rules with bra and ket determinants, task 0: with alpha and beta excitations, task 1: with one or two beta excitations and task 2: with one or two alpha excitations. The following code shows task 0 with alpha and beta excitations.

\begin{lstlisting}[float, language=c++,breaklines=true,columns=fixed,basewidth=0.5em,caption=task loop with alpha and beta excitations,label=lst:task loop]
for(ia = 0; ia < braAlphaSize; ia++){
  for(ib = 0; ib < braBetaSize; ib++){
    size_t braIdx = ia*braBetaSize + ib;
        
    DetI = DetFromAlphaBeta(adets[ia], bdets[ib]);
        
    for(j=0; j < SinglesFromAlphaLen[ia]; j++){
      size_t ja = SinglesFromAlphaSM[ia][j];
      for(k=0; k < SinglesFromBetaLen[ib]; k++){
        size_t jb = SinglesFromBetaSM[ib][k];
        size_t ketIdx = ja*ketBetaSize + jb;
        DetJ = DetFromAlphaBeta(adets[ja], bdets[jb]);
        ElemT eij = Hij(DetI,DetJ);
        Wb[braIdx] += eij * T[ketIdx];
      }
    }
  }
}
\end{lstlisting}

In this task, there are double nested loops for alpha and beta for each determinant. In the original implementation running on CPUs, only the topmost alpha loop is parallelized by OpenMP directive. To run on the GPUs we need as much threads as possible, so we collapse alpha and beta loops to parallelize. However, third and fourth loops have sparsity that causes load imbalances between GPU threads. We can collapse all the four loops and map them to GPU threads, but there will be data race to add multiplication results to \verb+Wb+ vector. We apply \verb+atomicAdd+ intrinsic that allows hardware accelerated atomic add operation on GPU to resolve data race.

\subsection{Determinant Caching}

There is another issue to parallelize task loops on the GPUs, bra and ket determinants for each alpha and beta are calculated for each GPU thread that requires thread local memory allocation for each GPU thread. GPU accept allocating dynamically on the fly in the GPU kernel codes but it is very slow. Also by collapsing four loops there will be huge number of threads and it will require very large memory space to store determinants for each threads that does not fit in the GPU memory. 

In the task loop code Listing \ref{lst:task loop}, \verb+DetI+ and \verb+DetJ+ are calculated for bra \verb+(ia, ib)+ and \verb+(ja, jb)+ alpha and beta for each thread. But this is redundant because the original implementation reuses \verb+DetI+ in third and fourth loops, and also some \verb+DetJ+ maybe redundant. To reduce redundancy and memory usage, we precalculate determinants for each \verb+(ia, ib)+ and \verb+(ja, jb)+ and caches them on the GPU memory. This determinant cache can be reused for the same range of alpha and beta for bra and ket defined in the task. The alpha and beta range of task varies when we distribute the diagonalization over the multiple processes by using MPI, so we recalculate determinant cache at the beginning of the task if the range is different from the previous task. The determinant cache requires \verb+alpha * beta * sizeof(determinant) * 2+ this requires much less memory size than allocating determinants for each thread for on-the-fly determinant calculation.

In summary, determinant caching significantly reduces redundant bitstring construction and lowers per-thread memory pressure, enabling scalable GPU execution.

\subsection{Object Based Implementation with Thrust}
The vector allocated on the GPU is stored as \verb+thrust::device_vector+, but we can not pass the vector to the GPU kernel code. There are two options to pass the data storage to the GPU kernel code, 1) pass the reference to each data element as a reference in kernel parameters, 2) store raw pointer to the vector as a member of the kernel class. The first case is safer implementation, but we have many data for multiplication and it will be complicated to handle all the data as parameter. So we store raw pointers in the kernel class as protected members so that they can not be accessed from outside of the kernel. 

\begin{figure}[!t] \centering \includegraphics[width=3in]{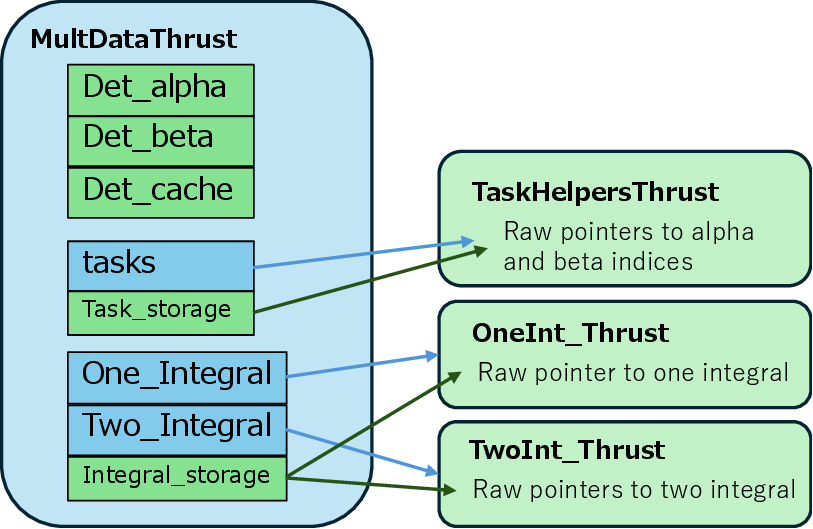} 
\caption{Data managements for Thrust based diagonalization. The MultDataThrust is the GPU data storage accessed from CPU side, and the raw pointers for each vector is extracted in the GPU kernel classes (shown in green boxes)}
\label{fig_data_structure} 
\end{figure}

Fig. \ref{fig_data_structure} shows data structure for the \verb+mult+ function. We define all the common data for the \verb+mult+ in the \verb+MultDataThrust+ class. This class is used from CPU side and holds vectors as \verb+thrust::device_vector+s. The \verb+TaskHelperThrust+ class has raw pointers to alpha and beta indices for the excitations. The vector itself is defined in the \verb+MultDataThrust+ class and the raw pointer is extracted at the construction.

\begin{figure}[!t] \centering \includegraphics[width=3in]{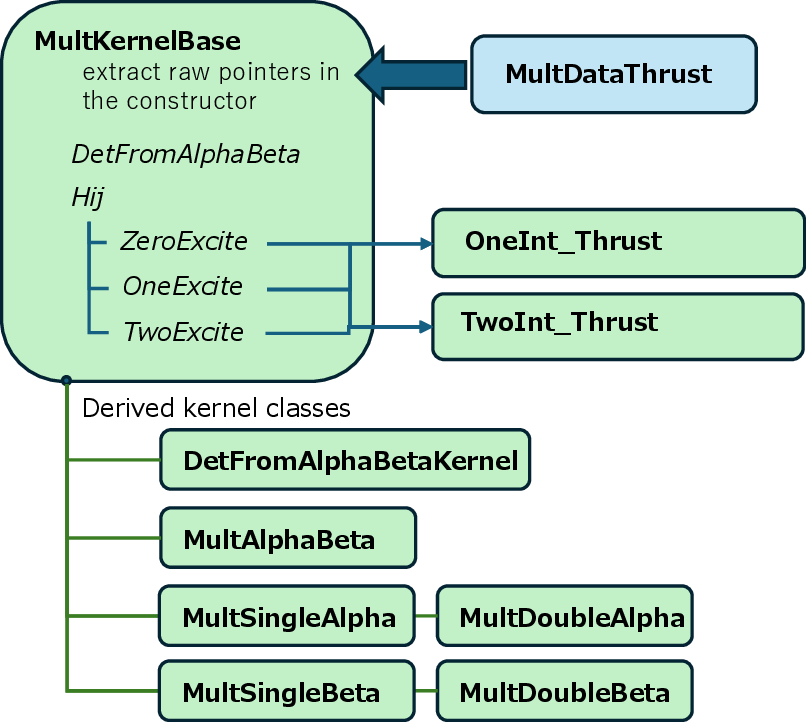} 
\caption{Core kernel classes for mult function. The MultKernelBase class is the base class with common functions used for every task types. The raw data is extracted at the construction from the MultDataThrust class.}
\label{fig_kernel_classes} 
\end{figure}

Fig. \ref{fig_kernel_classes} shows GPU kernel classes used in \verb+mult+ function. The \verb+MultKernelBase+ class is the base class containing common kernel functions and data. The data and raw pointers are set from \verb+MultDataThrust+ class in the constructor of the \verb+MultKernelBase+ class. The \verb+DetFromAlphaBeta+ function calculates determinants from alpha and beta coordinate used to precalculate determinant caches and the \verb+Hij+ function calculates Hamiltonian with bra and ket determinants. Most of the kernel codes in this class can be implemented by copying the original CPU codes into each class function. 

Then we can implement kernels for each task by deriving the \verb+MultKernelBase+ class, the \verb+MultAlphaBeta+ class for task type 0, the \verb+MultSingleBeta+ and the \verb+MultDoubleBeta+ class for task type 1 and the \verb+MultSingleAlpha+ and the \verb+MultDoubleAlpha+ class for task type 2.

This design ensures a clear separation between data management and kernel execution, improving maintainability and portability across GPU architectures.

\subsection{Overlapping Data Exchange and Calculations}

\begin{figure}[!t] \centering \includegraphics[width=3.5in]{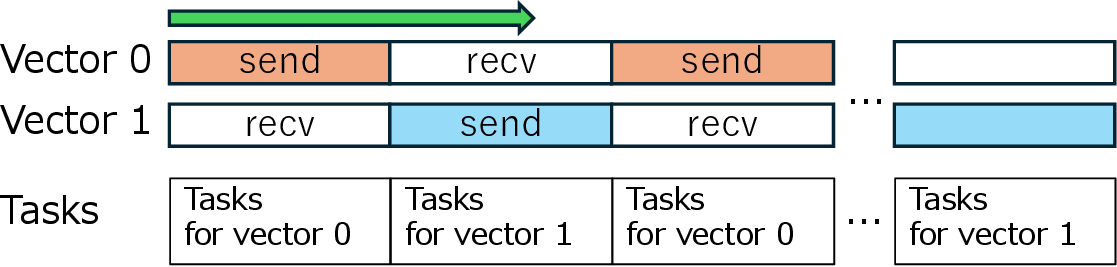} 
\caption{Overlapping data exchange between processes and task calculations by using double buffering technique. One of the two vectors is referred by tasks and also sent to another process while receiving data from another process in another vector. }
\label{fig_overlapping_comm} 
\end{figure}

To hide the data transfer time, we overlap the data transfer and the calculations by applying the double buffering technique. Fig. \ref{fig_overlapping_comm} shows the transition of two vectors and tasks for multiplications. At the first, the local vector stored in vector 0 is referred in each task calculation and vector 0 is also sent to other process asynchronously at the same time. The vector elements are sent to the destination directly from GPU memory by using RDMA. After calculating all the task for the local vector, we synchronize the data transfer to receive data in vector 1. Then we calculate tasks by referring to vector 1 and send to the next destination asynchronously. We repeat these cycles at the end of the tasks.

\subsection{Extension to Full-Bitstring Representations}

Although the present implementation focuses on half-bitstring determinants, the GPU kernel design naturally generalizes to full-bitstring representations. In the full-bitstring case, each configuration is stored as a single packed bitstring, and all single- and double-excitation patterns are generated via bitwise operations such as population count, bit scans, and parity evaluation. The collapsed-loop parallelization and \verb+atomicAdd+-based accumulation used in the half-bitstring kernels apply without modification; only the excitation generator and determinant cache need to be replaced with bitwise versions. Because the current data layout already stores determinants in flattened device-resident buffers and the kernel classes operate on raw pointers, enabling full-bitstring SBD primarily requires adding these new bitstring-level iterators rather than redesigning the overall execution pipeline. Full-bitstring support is under active development.

\section{Performance Evaluation}
We tested and evaluated the SBD code on the Miyabi supercomputer system for the Joint Center for Advanced High Performance Computing (JCAHPC) operated by the University of Tokyo and University of Tsukuba. The Miyabi supercomputer system consists of two clusters Miyabi-G and Miyabi-C. The Miyabi-G is the GPU cluster with 1,120 compute nodes of NVIDIA's GH200 Grace-Hopper (Table \ref{table_eval_env}). The Miyabi-C is Intel's Xeon cluster. We have run the SBD code on the Miyabi-G cluster. 

\begin{table}[!t] \renewcommand{\arraystretch}{1.3} \caption{Miyabi-G cluster overview} \label{table_eval_env} \centering 
\begin{tabular}{l|l}
\hline 
\hline 
    Compute node & NVIDIA GH200 \\\hline 
    CPU & NVIDIA Grace (72 cores)\\\hline 
    CPU memory size & 120 GB \\\hline 
    GPU & NVIDIA H200 \\\hline 
    GPU memory size & 96 GB \\\hline 
    Total number of nodes & 1,120 \\\hline 
    Total theoretical performance & 78.8 PFLOPS \\\hline 
    Interconnect & Infiniband NDR \\\hline 
\end{tabular} \end{table}

To demonstrate applicability to chemically relevant systems, we evaluate the proposed GPU-accelerated SBD implementation on the Fe$_4$S$_4$ cluster, using 6,000 quantum-sampled 36-bit configurations. This dataset results in $3.6\times 10^{7}$ determinants, corresponding to a representative SQD workload. The determinant cache for this problem requires approximately $2 \times 32 \times 3.6\times 10^{7}$ bytes ($\sim 2.3$ GB), allowing the entire Davidson iteration—including excitation lists, determinant data, and intermediate vectors—to fit comfortably within the 96~GB GPU memory of the NVIDIA H200.

We performed experiments on the Miyabi-G cluster of the JCAHPC facility (Table~\ref{table_eval_env}), which consists of 1,120 nodes equipped with NVIDIA GH200 Grace–Hopper processors. Each node integrates a 72-core Grace CPU with 120~GB LPDDR5X memory and a Hopper H200 GPU with 96~GB HBM3 memory, interconnected through NVLink-C2C and NDR Infiniband.



\begin{figure}[!t] \centering \includegraphics[width=3.5in]{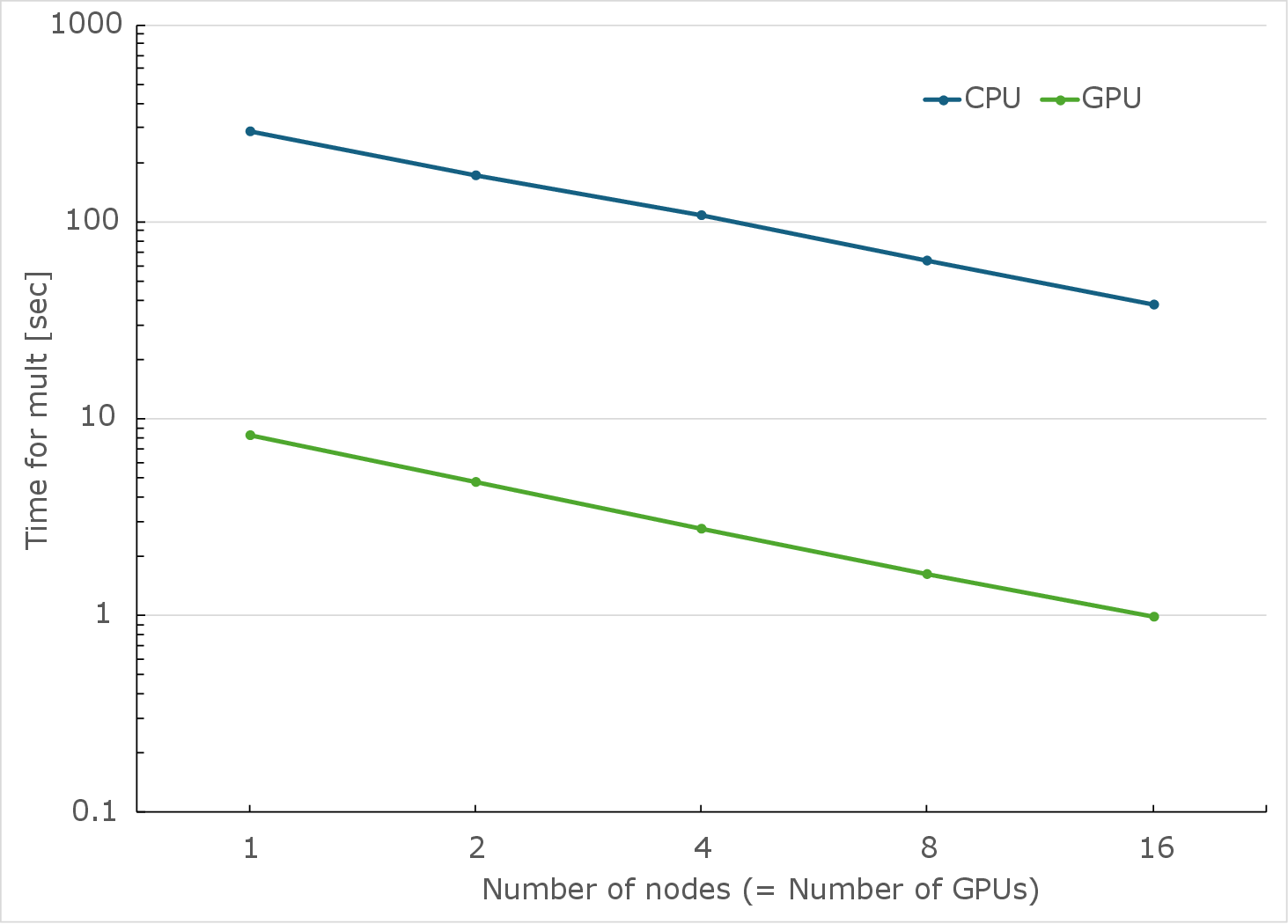} 
\caption{Elapsed time comparison of mult function with Grace CPU and H200 GPU on Miyabi-G cluster.}
\label{fig_scaling} 
\end{figure}

Figure~\ref{fig_scaling} summarizes the strong-scaling behavior of the \texttt{mult} kernel within Davidson, comparing Grace CPU and H200 GPU executions for $p\in\{1,2,4,8,16\}$ nodes. 
The average elapsed time per \texttt{mult} decreases from $288.05$~s to $37.68$~s on CPUs and from $8.25$~s to $0.99$~s on GPUs as $p$ increases from $1$ to $16$, showing good strong scaling on both backends. 
The per-node speedup $S_p = t^{\mathrm{CPU}}_p / t^{\mathrm{GPU}}_p$ remains essentially flat across node counts, ranging from $34.90\times$ (1 node) to $39.07\times$ (4 nodes) and $39.05\times$ (8 nodes), with a slight taper to $37.93\times$ at 16 nodes. 
This stability indicates that the GPU acceleration applies uniformly as the problem is decomposed, and that communication overheads affect both backends in a comparable way.

To quantify strong-scaling efficiency, we compute $E_p = t^{\mathrm{ref}}_1 / (p\, t^{\mathrm{ref}}_p)$ for each backend.
At $p=16$, the CPU path achieves $E_{16}^{\mathrm{CPU}}\approx 0.48$, while the GPU path reaches $E_{16}^{\mathrm{GPU}}\approx 0.52$. 
The modest drop in speedup at 16 nodes is consistent with increased synchronization and data exchange at larger $p$, even though we overlap MPI transfers with computation via double buffering. 
Notably, the GPU time per \texttt{mult} falls to $\sim$1~s at 16 nodes, bringing the cost of the dominant kernel to a regime where the overall SQD iteration time is primarily determined by the number of Davidson steps rather than single-kernel latency.

Overall, the results demonstrate (i) near-constant per-node speedup of $\sim$35--39$\times$ from 1 to 16 nodes, (ii) comparable or slightly better strong-scaling efficiency on GPUs relative to CPUs, and (iii) practical end-to-end acceleration that enables chemically relevant SQD workloads (e.g., Fe$_4$S$_4$) to complete within the memory and runtime budgets of modern GPU nodes.

\section{Conclusion}

We presented a GPU-accelerated implementation of Selected Basis Diagonalization tailored for SQD-based quantum–classical workflows. 
By restructuring the matrix-free Hamiltonian application around data-parallel execution, adopting the Thrust C++ library for GPU-resident vector operations, and introducing determinant caching to reduce redundant bitstring construction, the proposed method exposes large amounts of thread-level parallelism and minimizes device-side memory pressure. 
These design choices allow the algorithm to scale efficiently on modern GPU architectures while maintaining compatibility with both half-bitstring and full-bitstring representations.

We further introduced a modular, object-based kernel organization in which data management and device execution are cleanly separated, enabling maintainable and portable GPU kernels across evolving architectures. 
For distributed execution, we overlap MPI communication with GPU computation using a double-buffering strategy, mitigating communication overheads as the number of nodes increases.

Our evaluation on the Miyabi-G GH200 cluster demonstrated more than 35--39$\times$ per-node speedup over Grace CPUs across 1 to 16 nodes.
For a chemically relevant Fe$_4$S$_4$ problem with $3.6\times10^{7}$ determinants, the entire Davidson iteration fits in the 96~GB of GPU memory, and the dominant \texttt{mult} kernel executes in under one
second at 16 nodes. 
These results show that GPU-native SBD can significantly reduce the classical computational bottleneck in SQD, bringing diagonalization times into a regime compatible with near-term quantum sampling rates.

Overall, this work demonstrates that Thrust-based GPU implementations provide an effective and portable foundation for accelerating SBD on heterogeneous HPC systems. 
As GPU-accelerated supercomputers continue to dominate the HPC landscape, such GPU-native diagonalization techniques will be essential for scaling SQD and related hybrid quantum–classical algorithms to increasingly complex chemical and materials systems.

\section*{Acknowledgment}

T. S. and S. Y. acknowledge that part of this work was funded by NEDO through Project JPNP20017. 
They were also supported by JSPS KAKENHI Grants Nos. JP21H04446 and JP22K03479, JST COI-NEXT (Grant No. JPMJPF2221), 
and MEXT’s Program for Promoting Research of the Supercomputer Fugaku (Grant No. MXP1020230411). 
Additional support was provided by the UTokyo Quantum Initiative and the RIKEN TRIP initiative (RIKEN Quantum).

\end{document}